\documentclass[twocolumn]{aastex631}

\usepackage{natbib}

\usepackage[disable,colorinlistoftodos]{todonotes}

\begin{document}

\title{Give Me a Few Hours: Exploring Short Timescales in Rubin Observatory Cadence Simulations}

\author[0000-0001-8018-5348]{Eric C. Bellm}
\affiliation{DIRAC Institute, Department of Astronomy, University of Washington, 3910 15th Avenue NE, Seattle, WA 98195, USA}
\affiliation{ecbellm@uw.edu}

\author[0000-0001-9947-6911]{Colin J. Burke}
\affiliation{Department of Astronomy, University of Illinois at Urbana-Champaign, 1002 W. Green Street, Urbana, IL 61801, USA}

\author[0000-0002-8262-2924]{Michael W. Coughlin}
\affil{School of Physics and Astronomy, University of Minnesota, Minneapolis, Minnesota 55455, USA}

\author[0000-0003-3768-7515]{Igor Andreoni}
\affil{Division of Physics, Mathematics and Astronomy, California Institute of Technology, Pasadena, CA 91125, USA}
\affil{Joint Space-Science Institute, University of Maryland, College Park, MD 20742, USA}
\affil{Department of Astronomy, University of Maryland, College Park, MD 20742, USA}
\affil{Astrophysics Science Division, NASA Goddard Space Flight Center, Mail Code 661, Greenbelt, MD 20771, USA}

\author[0000-0003-1784-2784]{Claudia M.\ Raiteri}
\affiliation{INAF, Osservatorio Astrofisico di Torino, via Osservatorio 20, I-10025 Pino Torinese, Italy}

\author[0000-0001-9297-7748]{Rosaria Bonito}
\affiliation{INAF, Osservatorio Astronomico di Palermo, Piazza del Parlamento, 1 90134, Palermo, Italy}

\begin{abstract}
The limiting temporal resolution of a time-domain survey in detecting transient behavior is set by the time between observations of the same sky area.
We analyze the distribution of visit separations for a range of Vera C.\ Rubin Observatory cadence simulations.
Current simulations are strongly peaked at the 22 minute visit pair separation and provide effectively no constraint on temporal evolution within the night.
This choice will necessarily prevent Rubin from discovering a wide range of astrophysical phenomena in time to trigger rapid followup.
We present a science-agnostic metric to supplement detailed simulations of fast-evolving transients and variables and suggest potential approaches for improving the range of timescales explored.
\end{abstract}


\listoftodos

\section{Introduction}

For time-domain surveys, the temporal pattern of observations, or cadence, determines which astrophysical phenomena can be observed.
Accordingly, survey designers choose cadences based on their scientific goals in conjunction with the constraints imposed by their instruments \citep[cf.][]{Bellm:16:Cadences}.

The Vera C.\ Rubin Observatory \citep{2019ApJ...873..111I} aims to conduct a Legacy Survey of Space and Time (LSST) that addresses four broad science pillars: probing Dark Energy and Dark Matter, taking an inventory of the Solar System, exploring the transient optical sky, and mapping the Milky Way. 
Given the scale of the survey and the breadth of its scientific aims \citep{abell2009lsst}, the project is conducting extensive simulations  \citep{2014SPIE.9150E..14C} of potential cadences and evaluating them according to community-supplied metrics, which provide scientifically-motivated scores for evaluating the relative performance of different cadence simulations.
\citet{tmp_Bianco:21:CadenceNotesOverview} provides an overview of this process, which makes use of the OpSim simulation framework \cite{2016SPIE.9910E..13D, 2016SPIE.9911E..25R}, a feature-based scheduler \citep{2019AJ....157..151N}, and the Metrics Analysis Framework  \citep[MAF;][]{2014SPIE.9149E..0BJ}.

As a supplement to metrics which treat specific classes of transients \citep[e.g.,][]{Andreoni:21:KilonovaCadence}, variables, and accreting sources (e.g., \citealp{Bonito2021}, Raiteri et al. in prep.), we present a source-agnostic analysis of the time gaps present in current cadence simulations.
While detailed metrics simulating specific object classes are important in determining the science impacts of specific cadence choices, they require extensive development by domain experts and may not span the discovery space.
Additionally it is challenging to weight specific object classes against one another.
Simple metrics in conjunction with knowledge of the survey design can provide a useful supplement to scientifically-motivated analyses.

The (logarithmic) time separation of visits to a given sky area encapsulates the most basic information content of a time-domain survey.
Our goal is to maximize the information we gain from these visits about time-varying objects.
As discussed by \citet{Richards2018}, for a sparsely-sampled time-domain survey,
an ideal cadence for source class-agnostic discovery and variability characterization would be uniform in $\log \Delta t$---it would be sensitive to variations on all timescales, from the length of a single exposure up to the total survey duration.

In practice, of course, a ground-based survey cannot achieve this uniformity due to diurnal and seasonal cycles.
However, the cadence families explored in LSST simulations at the time of this writing \citep{jones_r_lynne_2020_4048838} are still far from effective at probing the full range of accessible timescales.
In particular, the use of closely-spaced observation pairs leaves an ``intra-night desert'' preventing real-time discovery and timely followup of any phenomena varying on timescales of a few hours to a day.
This includes stellar flares; young supernovae; gamma-ray bursts, orphan afterglows, and other relativistic transients; kilonovae; and rare new kinds of fast extragalactic transients.
This also includes a variety short-timescale accretion variability: 
Young Stellar Objects (YSOs) show short-timescale bursts and dipping events due to a variety of physical mechanisms, including accretion rate changes, disk warping, stellar flares, and starspots \citep[e.g.,][]{Bonito:18:YSOLSST}.
Among Active Galactic Nuclei (AGN), low-mass supermassive black holes and accreting intermediate mass black holes are expected to be most variable on hours to days timescales \citep[e.g.,][]{2021Sci...373..789B}.
Likewise, the most dramatic active phases of extreme flaring blazars exhibit short-timescale variability that can help identify the underlying emission physics \citep[e.g.,][]{Raiteri:21a, Raiteri:21b}.

In this paper we present a new scalar metric for evaluating the temporal sampling of a time-domain survey on timescales of interest (\S \ref{sec:metric}).
In \S \ref{sec:results} we evaluate this metric on current LSST cadence simulations, with particular focus on sampling at short timescales.
We close in \S \ref{sec:discussion} with a range of ideas for diversifying the cadences of the Rubin Observatory's LSST.

\section{Metrics} \label{sec:metric}

To analyze the current simulations, we used the existing MAF metric \texttt{lsst.sims.maf.metrics.TgapsMetric} with logarithmic bins from 30 seconds to 10 years.\footnote{\texttt{np.logspace(-3.46,3.54,99)}; see \url{https://github.com/RichardsGroup/LSST_OpSim/blob/master/contrib/00_computeLogTgapsMetric.ipynb}}.
We computed pairwise separations (\texttt{allGaps=False}) between observations in any filter and summed the resulting histograms for an NSIDE=64 healpix grid\footnote{This restriction to gaps between consecutive observations understates the temporal information present in more densely sampled surveys with many observations within a period less than the astrophysical timescales of interest.  For instance, these histograms for a single sector of the Transiting Exoplanet Survey Satellite \citep[TESS;][]{Ricker:14:TESS} would consist of a single spike at the continuous 30-minute full-frame image cadence, but the data would provide temporal information over the entire 27-day interval in which TESS was pointed at the sector.}.
We also computed a normalized cumulative distribution function (CDF) of the resulting histogram. 

Using this cumulative histogram as a conceptual starting point, we defined a new metric (\texttt{TgapsPercentMetric}) which represents the percentage of observation pairs with separations between 2 hours and 14 hours (same night revisits) and between 14 and 38 hours (next-night revisits).
We selected these intervals due to their importance for identifying fast-evolving transients, but the metric can be configured to use any desired time window.
The resulting code is publicly available in the central \texttt{rubin\_sim} MAF metric repository\footnote{\url{https://github.com/lsst/rubin\_sim/blob/main/rubin\_sim/maf/metrics/tgaps.py}}.

\begin{figure}
    \centering
    \includegraphics[width=\columnwidth]{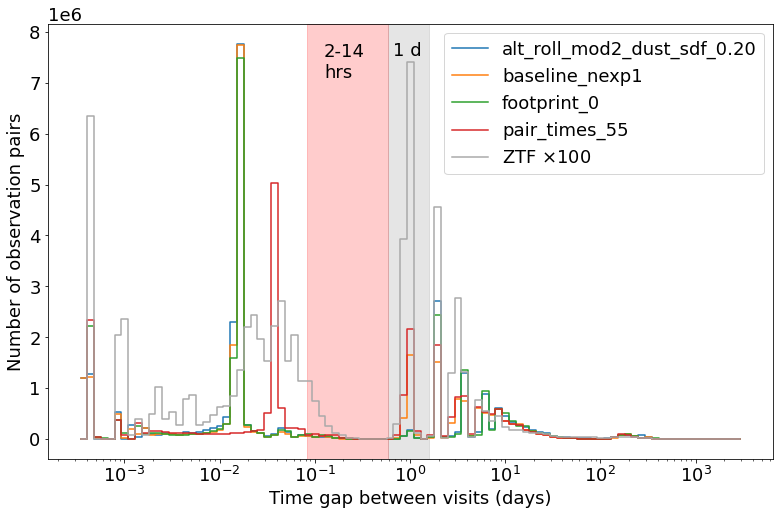}
    \includegraphics[width=\columnwidth]{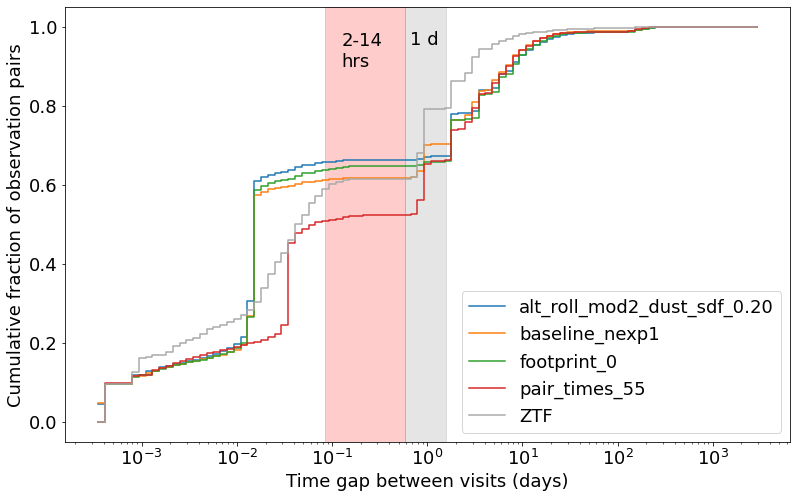}
    \caption{Histogram (top) and normalized cumulative histogram (bottom) of the time gaps between visits to the same sky position for several representative LSST simulations as well as Zwicky Transient Facility (ZTF) observations.
    Shaded regions highlight the 2--14 hour and 14--38 hour ranges for which we compute metrics (Table \ref{tab:opsim}).  
    \label{fig:dt_hist}}
\end{figure}

\section{Results} \label{sec:results}

We evaluated our metrics on \texttt{OpSim} simulations from the v1.5, v1.7, and v1.7.1 families \citep{jones_r_lynne_2020_4048838}.
Figure \ref{fig:dt_hist} presents these time gap histograms for several illustrative example runs, and Table \ref{tab:opsim} summarizes the \texttt{TgapsPercentMetric} for each.

Despite the possibility of observing fields with time gaps longer than two hours but before the end of the current night, we find that less than 1\% of visits are spaced in this critical timescale across all survey families.
We stress that there is no inherent limitation preventing observations in this time range.

Perhaps surprisingly, given the fiducial 3-day cadence of the main Wide-Fast-Deep survey, one-night cadence timescales are somewhat better covered. 
Many simulation families show 7--12\% of visit gaps at one-day timescales, although a subset have sub-percent fractions at 1-day timescales as well.
The latter observing strategies would be catastrophic for discovery of fast-evolving transients.

We also compared these simulations to comparable histograms from the Zwicky Transient Facility \citep[ZTF;][]{Bellm:19:ZTFOverview, Graham:19:ZTFScience}.
While ZTF has a field of view five times larger than LSSTCam, and therefore on average revisits a given area of sky five times more often \citep{Bellm:16:Cadences}, the ZTF surveys have also explicitly sought to span a wide range of timescales.
These have included ``movie mode'' continuous-cadence observations, 6 visit per night transient searches, twilight searches for moving objects, as well as slower 2, 3, and 4-day cadence surveys \citep{Bellm:19:ZTFScheduler}.

\begin{table*}[ht]
    \centering
    \begin{tabular}{|c|c|c|}
    \hline
    OpSim run & \texttt{TgapsPercentMetric} & \texttt{TgapsPercentMetric} \\
    & 2--14 hours & 14--38 hours \\
    \hline
    baseline\_nexp1 & \textit{0.5\%} & 8.6\% \\
    baseline\_nexp2 & \textit{0.5\%} & 8.2\% \\
    \hline
third\_obs\_pt15 & \textit{0.3\%} & 8.5\% \\
third\_obs\_pt30 & \textit{0.4\%} & 8.2\% \\
third\_obs\_pt45 & \textit{0.4\%} & 8.1\% \\
third\_obs\_pt60 & \textit{0.5\%} & 8.0\% \\
third\_obs\_pt90 & \textit{0.7\%} & 7.6\% \\
    third\_obs\_pt120 & \textit{0.8\%} & 7.3\% \\
    \hline
    pair\_times\_11 & \textit{0.4\%} & 7.5\% \\
    pair\_times\_55 & 1.3\% & \textbf{13.9\%} \\
    \hline
    
\hline
wfd\_depth\_scale0.65 & \textit{0.3\%} & 7.9\% \\
wfd\_depth\_scale0.99 & \textit{0.2\%} & 8.6\% \\
\hline
footprint\_0 & \textit{0.6\%} & \textit{0.2\%} \\
footprint\_stuck\_rolling & \textit{0.4\%} & 7.1\% \\
footprint\_big\_sky & \textit{0.2\%} & 8.8\% \\
filterdist\_indx1 & \textit{0.3\%} & 9.2\% \\
\hline
rolling\_scale0.2\_nslice2 & \textit{0.8\%} & \textit{0.1\%} \\
rolling\_scale1.0\_nslice3 & \textit{0.8\%} & \textit{0.1\%} \\
alt\_roll\_mod2\_dust\_sdf\_0.2 & \textit{0.5\%} & \textit{0.6\%} \\
roll\_mod2\_dust\_sdf\_0.20 & \textit{0.3\%} & 8.4\% \\
rolling\_nm\_scale1.0\_nslice2 & \textit{0.6\%} & 8.9\% \\
rolling\_nm\_scale0.90\_nslice3\_fpw0.9\_nrw1.0 & \textit{0.8\%} & \textbf{12.6\%} \\
\hline
\hline
Zwicky Transient Facility (observed) & 3.6\% & 17.7\% \\
\hline
    \end{tabular}
    \caption{Median percent of observations probing intra-night and 1-day timescales.
    We selected representative examples from within the simulation families for brevity.
    Metric values above 10\% are bolded, while sub-percent values are italicized.
    As-observed values for the Zwicky Transient Facility are included for comparison.
    }
    \label{tab:opsim}
\end{table*}

\subsection{Survey Footprint}

Generally, we expected the increased number of visits in simulations which use a smaller main survey (Wide-Fast-Deep) footprint would provide more effective time sampling.  
However, current simulation families do not distribute the additional visits at short timescales and so do not provide a major improvement in the relevant metrics.
The \texttt{wfd\_depth}, \texttt{filt\_dist}, and \texttt{footprint} simulations are comparable to or slightly worse than the current \texttt{baseline} simulations, with the exception of the \texttt{fbs\_1.7\/footprint\_tune\/footprint\_\#} simulations, which have extremely low (sub-percent) coverage at 1-day timescales. 

\subsection{Visit Pairs}

Paired observations need to be closely spaced (up to $\sim$ one hour) for linking of main-belt asteroids to succeed \citep{2018Icar..303..181J} due to the $N^2$ combinatoric explosion of source pairs to consider when constructing tracklets.  
Current simulation families explore the effect of varying both the visit pair separation as well as the filters used in the visit pair.

The \texttt{pair\_times} family considered spaces as large as 55 minutes.  
The largest pair-spacing simulation, \texttt{pair\_times\_55}, provides the best metric values for both the 2--14 and 14--38 hour timescales of all the simulations to date.  

With the exception of stellar flares, most of the classes of  fast transients motivating this work above will not vary appreciably on timescales less than an hour.
Accordingly this suggests that  switching filters between observations in a pair (as in the current baseline) is preferable, so that the second observation provides non-redundant information (color). 

\subsection{Triplet Observations}

``Triplet'' observations (the ``Presto-Color'' strategy,  \citealp{Bianco:19:PrestoColor}) add a third nightly observation in one of the two filters of the visit pair, and so present the best means of capturing the intranight variability of fast transients.
Suprisingly, the current \texttt{third\_obs\_pt\#} simulations show almost \textit{no} improvement in 2--14 hour timescale coverage. 
The current triplet implementation thus provides little improvement over the baseline strategy, as it does not provide sufficiently wide time sampling nor enough additional visits to substantially change the fraction of observations in the intra-night desert.
Further improvements to the current LSST implementation of this survey strategy are necessary for triplet observations to reach their potential.
Scheduling multiple images with wide spacing during the night may benefit from scheduling algorithms that optimize field selection on nightly timescales \citep[e.g.,][]{Bellm:19:ZTFScheduler}.

\subsection{Rolling Cadences}

Rolling cadences provide the best means of allocating additional observations into the 2 hours--1 night window critical for rapid discovery of fast-evolving transients.
Current simulations show a wide range of performance; the \texttt{rolling\_scale\*} and \texttt{alt\_roll} simulations have very poor (sub-percent) coverage of one-day timescales. 
\texttt{rolling\_nm\_scale1.0\_nslice2} is close to the \texttt{baseline}, and 
\texttt{rolling\_nm\_scale0.90\_nslice3\_fpw0.9\_nrw1.0} approaches the \texttt{pair\_times\_55} simulation in its effective timescale coverage.
We suggest continued investigation into how best to distribute the rolling visits in time.

\subsection{Per-filter Time Gaps}

\begin{figure}
    \centering
    \includegraphics[width=\columnwidth]{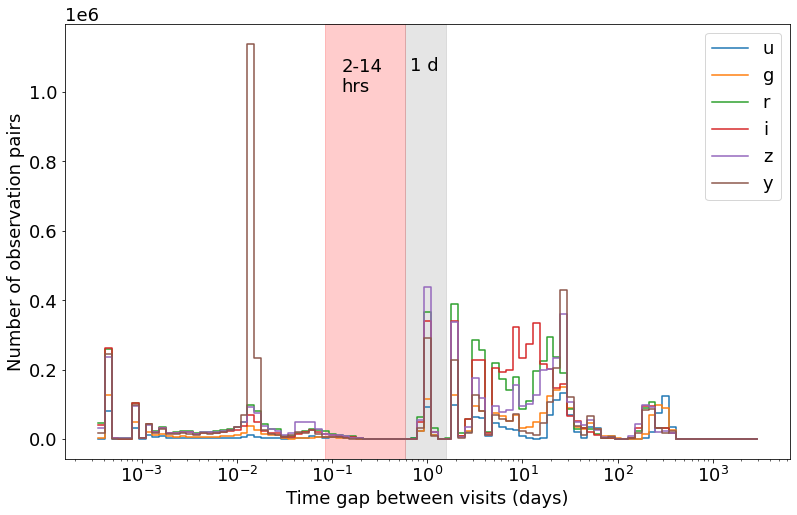}
    \includegraphics[width=\columnwidth]{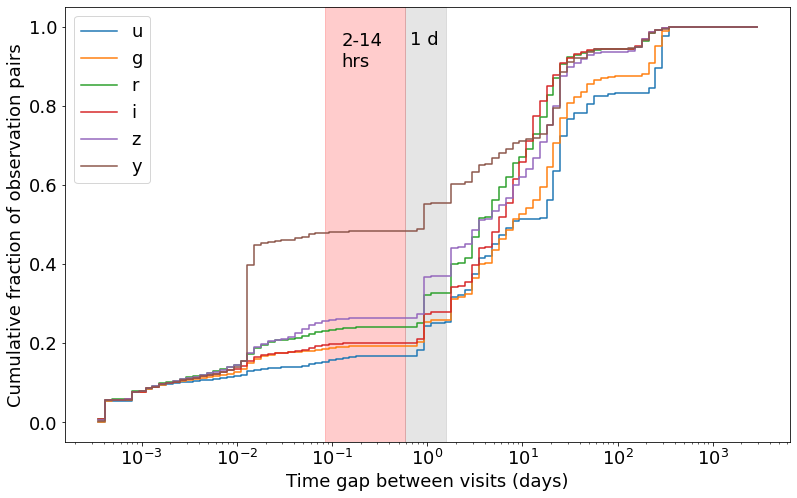}
    \caption{Histogram (top) and normalized cumulative histogram (bottom) of the time gaps between visits to the same sky position in the same filter in the \texttt{baseline\_nexp1} simulation excluding the Deep Drilling Fields.
    Shaded regions as in Figure \ref{fig:dt_hist}.
    \label{fig:dt_hist_byband}}
\end{figure}

So far we have considered pairwise time gaps between observations in any pair of filters.  
Analysis of short-timescale variability will be more straightforward when those two visits are in the same filter, however.
Figure \ref{fig:dt_hist_byband} shows the per-band time gap histograms between observations in the same filter for the \texttt{baseline\_nexp1} simulation.
We excluded the Deep Drilling Fields, which observe only a small sky area and cycle through the available filter set on a daily basis.

Overall, the morphology of the per-band time-gap histograms is quite similar to the total histograms (Figure \ref{fig:dt_hist}).
There are very few observation pairs in the 2--14 hours range (\texttt{TgapsPercentMetric} ranges from 0.0\% ($g$ \& $y$ bands) to 1.6\% ($u$)).
The 14--38 hour interval shows a larger revisit fraction, with \texttt{TgapsPercentMetric} ranging from 6.1\% ($g$) to 10.2\% ($z$).
The per-band sampling on one-day timescales is thus larger than expected given the notional baseline cadence, although we note that the histograms show a long tail of revisit times extending from weeks to months.

\subsection{Comparison to ZTF}

For an informative comparison to an existing sky survey, we also computed histograms and the \texttt{TgapsPercentMetric} on ZTF's on-sky pointing history.
We included pointings from all public and private surveys from March 2018--September 2021.
We determined pairwise time gaps for each discrete ZTF field, neglecting the overlaps at the field edges, which typically provide additional sampling on sub-hour timescales.  
Figure \ref{fig:dt_hist} shows the resulting time gap histogram, and Table \ref{tab:opsim} lists the corresponding metrics.

As discussed in \citet{Bellm:16:Cadences}, the areal survey rate (the instantaneous field of view of a survey camera divided by its exposure time and any overheads) determines the number of exposures of a field a survey can obtain on average.
Due to its wider field of view, ZTF obtains a factor of five more yearly exposures per field than LSST.
Despite this advantage and a significant focus of a subset of the private ZTF surveys on high-cadence observations \citep{Bellm:19:ZTFScheduler}, we see that ZTF has only 3.6\% of its observations in the 2--14 hour window and 17.7\% in the 14-38 hour window.
This is due to the presence of other large observing programs (e.g., the public 2- and 3-day cadence Northern Sky Survey).
We may thus take the ZTF numbers as a practical upper bound which LSST can achieve with its multi-faceted survey goals, and suggest that LSST target 1--2\% for 2--14 hours and 10--15\% for the 14--38 hour window.

\section{Discussion} \label{sec:discussion}

The existing survey simulations analyzed in \S \ref{sec:results} generally exhibit sub-percent revisit fractions within the night (Table \ref{tab:opsim}).
Modifications of the existing triplet and rolling cadence strategies may already be enough to improve this sampling.
However, we also suggest exploration of more unique cadence modes not present in the existing survey families.

Asteroid discovery drives the requirement for visit pairs spaced by an hour or less.
However, most Main Belt Asteroids ($\sim80\%$; M.\ Juric, priv.\ comm) are discovered in the first three years of the survey \citep{Ridgway:14:TransientDiscoveryRate}. 
Accordingly a move to much wider visit pairs ($>2$ hours) later in the survey might enhance fast transient discovery without compromising LSST's solar system science: with the majority of new asteroids discovered and the false-positive rate well-understood, identifying tracklets over wider temporal spacing could be tractable.
Alternatively, new asteroid discovery algorithms show promise in discovering asteroids independent of the input cadence \citep{tmp_Moeyens:21:THOR}.

Because the current LSST Deep Drilling Fields are scheduled as single contiguous blocks of observations, they also do not provide leverage for intra-night variability.
Scheduling approaches that separated observations of a Deep Drilling Field widely within the night would be extremely valuable for identifying short-timescale variability, albeit over a limited sky area.


Throughout this work we have largely focused on pairwise time separations between observations in any pair of filters.  
In practice, identifying rapid short-timescale variability---especially in near real-time---is most straightforward if those observations are both taken in the same filter.
Making use of heterogeneous filters will require model-dependent assumptions about the source spectral energy distribution and extinction. 
Although this challenge will be present in analysis of LSST's multi-band data at all timescales, the limited number of data points available for fast transients will make such interpretation particularly difficult.
For this reason, the Presto-Color triplet strategy \citep{Bianco:19:PrestoColor} explicitly requests two widely-spaced visits within the night in the same filter.

Were additional visits available, we would suggest the observing time be used to provide a ``variability wedding cake'' survey approach that would more broadly explore the discovery space in $\log \Delta t$.
This might include point-and-stare continuous-cadence (``movie-mode'') observations of single fields to provide sensitivity to very short timescale variability; short high-cadence campaigns on a limited sky area \citep[e.g.,][]{Bonito:18:YSOLSST}; adding third (triplet) observations within the night to Wide-Fast-Deep fields; and maximizing the season length to provide sampling on many month timescales.
For efficiency these additional observations might only use a subset of the available filters.
Reserving a few percent of observing time for a series of such experiments throughout the survey might yield outsized scientific returns.

The Vera C.\ Rubin Observatory's Legacy Survey of Space and Time will transform time-domain astronomy over the next decade.
Fulfilling its discovery potential across a wide range of science requires challenging cadence optimization.
In particular, efficient asteroid discovery requires closely-spaced visit pairs early in the survey.
The moderate number of available visits per year ($\sim$80, on average) to a given sky position and the need for effective sampling of supernovae on timescales of months further limit the ease with which LSST can observe short variability timescales.
Nevertheless, we argue that LSST should consider creative opportunities to enable exploration of the widest range of time-domain astrophysics.


\begin{acknowledgments}

We thank Lynne Jones, Peter Yoachim, and Mario Juric for useful discussions.

This paper was created in the Rubin LSST Transient and Variable Star (TVS) Science Collaboration \footnote{\url{https://lsst-tvssc.github.io/}}. The authors acknowledge the support of the Vera C.\ Rubin Legacy Survey of Space and Time Transient and Variable Stars Science Collaboration that provided opportunities for collaboration and exchange of ideas and knowledge and of Rubin Observatory in the creation and implementation of this work.
Publication support was provided by the Heising-Simons Foundation through the ``Preparing for Astrophysics with LSST'' grant to TVS and other LSST Science Collaborations.
\todo{tweak HS publication acknowledgement}
The authors also acknowledge the support of the LSST Corporation, which enabled the organization of workshops and hackathons throughout the cadence optimization process by directing private funding to these activities.

This research uses services or data provided by the Astro Data Lab at NSF's National Optical-Infrared Astronomy Research Laboratory. NOIRLab is operated by the Association of Universities for Research in Astronomy (AURA), Inc. under a cooperative agreement with the National Science Foundation.

ECB gratefully acknowledges support from the NSF AAG grant 1812779 and grant \#2018-0908 from the Heising-Simons Foundation.

ECB acknowledges further support from the Vera C.\ Rubin Observatory, which is supported in part by the National Science Foundation through
Cooperative Agreement 1258333 managed by the Association of Universities for Research in Astronomy
(AURA), and the Department of Energy under Contract No. DE-AC02-76SF00515 with the SLAC National
Accelerator Laboratory. Additional LSST funding comes from private donations, grants to universities,
and in-kind support from LSSTC Institutional Members.

\software{LSST Metrics Analysis Framework \citep[MAF;][]{2014SPIE.9149E..0BJ},
\texttt{Astropy} \citep{astropy:2013, astropy:2018},
\texttt{Numpy} \citep{numpy:2011, harris2020array},
\texttt{Matplotlib} \citep{Hunter:2007},
\texttt{healpy} \citep{2005ApJ...622..759G, Zonca2019}.}

\end{acknowledgments}

\bibliographystyle{aasjournal}
\bibliography{references,LSSTbiblio}

\begin{thebibliography}{}
\expandafter\ifx\csname natexlab\endcsname\relax\def\natexlab#1{#1}\fi
\providecommand{\url}[1]{\href{#1}{#1}}
\providecommand{\dodoi}[1]{doi:~\href{http://doi.org/#1}{\nolinkurl{#1}}}
\providecommand{\doeprint}[1]{\href{http://ascl.net/#1}{\nolinkurl{http://ascl.net/#1}}}
\providecommand{\doarXiv}[1]{\href{https://arxiv.org/abs/#1}{\nolinkurl{https://arxiv.org/abs/#1}}}

\bibitem[{Abell {et~al.}(2009)Abell, Allison, Anderson, Andrew, Angel, Armus,
  Arnett, Asztalos, Axelrod, Bailey, {et~al.}}]{abell2009lsst}
Abell, P.~A., Allison, J., Anderson, S.~F., {et~al.} 2009.
\newblock \doarXiv{0912.0201}

\bibitem[{{Andreoni} {et~al.}(2021){Andreoni}, {Coughlin}, {Almualla}, {Bellm},
  {Bianco}, {Bulla}, {Cucchiara}, {Dietrich}, {Goobar}, {Kool}, {Li},
  {Ragosta}, {Sagues-Carracedo}, \& {Singer}}]{Andreoni:21:KilonovaCadence}
{Andreoni}, I., {Coughlin}, M.~W., {Almualla}, M., {et~al.} 2021, arXiv
  e-prints, arXiv:2106.06820.
\newblock \doarXiv{2106.06820}

\bibitem[{{Astropy Collaboration} {et~al.}(2013){Astropy Collaboration},
  {Robitaille}, {Tollerud}, {Greenfield}, {Droettboom}, {Bray}, {Aldcroft},
  {Davis}, {Ginsburg}, {Price-Whelan}, {Kerzendorf}, {Conley}, {Crighton},
  {Barbary}, {Muna}, {Ferguson}, {Grollier}, {Parikh}, {Nair}, {Unther},
  {Deil}, {Woillez}, {Conseil}, {Kramer}, {Turner}, {Singer}, {Fox}, {Weaver},
  {Zabalza}, {Edwards}, {Azalee Bostroem}, {Burke}, {Casey}, {Crawford},
  {Dencheva}, {Ely}, {Jenness}, {Labrie}, {Lim}, {Pierfederici}, {Pontzen},
  {Ptak}, {Refsdal}, {Servillat}, \& {Streicher}}]{astropy:2013}
{Astropy Collaboration}, {Robitaille}, T.~P., {Tollerud}, E.~J., {et~al.} 2013,
  \aap, 558, A33, \dodoi{10.1051/0004-6361/201322068}

\bibitem[{{Astropy Collaboration} {et~al.}(2018){Astropy Collaboration},
  {Price-Whelan}, {Sip{\H{o}}cz}, {G{\"u}nther}, {Lim}, {Crawford}, {Conseil},
  {Shupe}, {Craig}, {Dencheva}, {Ginsburg}, {Vand erPlas}, {Bradley},
  {P{\'e}rez-Su{\'a}rez}, {de Val-Borro}, {Aldcroft}, {Cruz}, {Robitaille},
  {Tollerud}, {Ardelean}, {Babej}, {Bach}, {Bachetti}, {Bakanov}, {Bamford},
  {Barentsen}, {Barmby}, {Baumbach}, {Berry}, {Biscani}, {Boquien}, {Bostroem},
  {Bouma}, {Brammer}, {Bray}, {Breytenbach}, {Buddelmeijer}, {Burke},
  {Calderone}, {Cano Rodr{\'\i}guez}, {Cara}, {Cardoso}, {Cheedella}, {Copin},
  {Corrales}, {Crichton}, {D'Avella}, {Deil}, {Depagne}, {Dietrich}, {Donath},
  {Droettboom}, {Earl}, {Erben}, {Fabbro}, {Ferreira}, {Finethy}, {Fox},
  {Garrison}, {Gibbons}, {Goldstein}, {Gommers}, {Greco}, {Greenfield},
  {Groener}, {Grollier}, {Hagen}, {Hirst}, {Homeier}, {Horton}, {Hosseinzadeh},
  {Hu}, {Hunkeler}, {Ivezi{\'c}}, {Jain}, {Jenness}, {Kanarek}, {Kendrew},
  {Kern}, {Kerzendorf}, {Khvalko}, {King}, {Kirkby}, {Kulkarni}, {Kumar},
  {Lee}, {Lenz}, {Littlefair}, {Ma}, {Macleod}, {Mastropietro}, {McCully},
  {Montagnac}, {Morris}, {Mueller}, {Mumford}, {Muna}, {Murphy}, {Nelson},
  {Nguyen}, {Ninan}, {N{\"o}the}, {Ogaz}, {Oh}, {Parejko}, {Parley}, {Pascual},
  {Patil}, {Patil}, {Plunkett}, {Prochaska}, {Rastogi}, {Reddy Janga},
  {Sabater}, {Sakurikar}, {Seifert}, {Sherbert}, {Sherwood-Taylor}, {Shih},
  {Sick}, {Silbiger}, {Singanamalla}, {Singer}, {Sladen}, {Sooley},
  {Sornarajah}, {Streicher}, {Teuben}, {Thomas}, {Tremblay}, {Turner},
  {Terr{\'o}n}, {van Kerkwijk}, {de la Vega}, {Watkins}, {Weaver}, {Whitmore},
  {Woillez}, {Zabalza}, \& {Astropy Contributors}}]{astropy:2018}
{Astropy Collaboration}, {Price-Whelan}, A.~M., {Sip{\H{o}}cz}, B.~M., {et~al.}
  2018, \aj, 156, 123, \dodoi{10.3847/1538-3881/aabc4f}

\bibitem[{{Bellm}(2016)}]{Bellm:16:Cadences}
{Bellm}, E.~C. 2016, \pasp, 128, 084501

\bibitem[{{Bellm} {et~al.}(2019{\natexlab{a}}){Bellm}, {Kulkarni}, {Graham},
  {Dekany}, {Smith}, {Riddle}, {Masci}, {Helou}, {Prince}, {Adams},
  {Barbarino}, {Barlow}, {Bauer}, {Beck}, {Belicki}, {Biswas}, {Blagorodnova},
  {Bodewits}, {Bolin}, {Brinnel}, {Brooke}, {Bue}, {Bulla}, {Burruss}, {Cenko},
  {Chang}, {Connolly}, {Coughlin}, {Cromer}, {Cunningham}, {De}, {Delacroix},
  {Desai}, {Duev}, {Eadie}, {Farnham}, {Feeney}, {Feindt}, {Flynn},
  {Franckowiak}, {Frederick}, {Fremling}, {Gal-Yam}, {Gezari}, {Giomi},
  {Goldstein}, {Golkhou}, {Goobar}, {Groom}, {Hacopians}, {Hale}, {Henning},
  {Ho}, {Hover}, {Howell}, {Hung}, {Huppenkothen}, {Imel}, {Ip}, {Ivezi{\'c}},
  {Jackson}, {Jones}, {Juric}, {Kasliwal}, {Kaspi}, {Kaye}, {Kelley},
  {Kowalski}, {Kramer}, {Kupfer}, {Landry}, {Laher}, {Lee}, {Lin}, {Lin},
  {Lunnan}, {Giomi}, {Mahabal}, {Mao}, {Miller}, {Monkewitz}, {Murphy},
  {Ngeow}, {Nordin}, {Nugent}, {Ofek}, {Patterson}, {Penprase}, {Porter},
  {Rauch}, {Rebbapragada}, {Reiley}, {Rigault}, {Rodriguez}, {van Roestel},
  {Rusholme}, {van Santen}, {Schulze}, {Shupe}, {Singer}, {Soumagnac}, {Stein},
  {Surace}, {Sollerman}, {Szkody}, {Taddia}, {Terek}, {Van Sistine}, {van
  Velzen}, {Vestrand}, {Walters}, {Ward}, {Ye}, {Yu}, {Yan}, \&
  {Zolkower}}]{Bellm:19:ZTFOverview}
{Bellm}, E.~C., {Kulkarni}, S.~R., {Graham}, M.~J., {et~al.}
  2019{\natexlab{a}}, \pasp, 131, 018002

\bibitem[{{Bellm} {et~al.}(2019{\natexlab{b}}){Bellm}, {Kulkarni}, {Barlow},
  {Feindt}, {Graham}, {Goobar}, {Kupfer}, {Ngeow}, {Nugent}, {Ofek}, {Prince},
  {Riddle}, {Walters}, \& {Ye}}]{Bellm:19:ZTFScheduler}
{Bellm}, E.~C., {Kulkarni}, S.~R., {Barlow}, T., {et~al.} 2019{\natexlab{b}},
  \pasp, 131, 068003

\bibitem[{{Bianco} {et~al.}(2019){Bianco}, {Drout}, {Graham}, {Pritchard},
  {Biswas}, {Narayan}, {Andreoni}, {Cowperthwaite}, {Ribeiro}, {LSST
  Transient}, \& {Variable Stars Collaboration}}]{Bianco:19:PrestoColor}
{Bianco}, F.~B., {Drout}, M.~R., {Graham}, M.~L., {et~al.} 2019, \pasp, 131,
  068002, \dodoi{10.1088/1538-3873/ab121a}

\bibitem[{{Bianco} {et~al.}(2021){Bianco}, {Ivezi{\'c}}, {Jones}, {Graham},
  {Marshall}, {Saha}, {Strauss}, {Yoachim}, {Ribeiro}, {Anguita}, {Bauer},
  {Bellm}, {Blum}, {Brandt}, {Brough}, {Catelan}, {Clarkson}, {Connolly},
  {Gawiser}, {Gizis}, {Hlozek}, {Kaviraj}, {Liu}, {Lochner}, {Mahabal},
  {Mandelbaum}, {McGehee}, {Neilsen}, {Olsen}, {Peiris}, {Rhodes}, {Richards},
  {Ridgway}, {Schwamb}, {Scolnic}, {Shemmer}, {Slater}, {Slosar}, {Smartt},
  {Strader}, {Street}, {Trilling}, {Verma}, {Vivas}, {Wechsler}, \&
  {Willman}}]{tmp_Bianco:21:CadenceNotesOverview}
{Bianco}, F.~B., {Ivezi{\'c}}, {\v{Z}}., {Jones}, R.~L., {et~al.} 2021, arXiv
  e-prints, arXiv:2108.01683.
\newblock \doarXiv{2108.01683}

\bibitem[{{Bonito} {et~al.}(2018){Bonito}, {Hartigan}, {Venuti}, {Guarcello},
  {Prisinzano}, {Argiroffi}, {Messina}, {Johns-Krull}, {Feigelson}, {Stauffer},
  {Giannini}, {Antoniucci}, {Sciortino}, {Micela}, {Pillitteri}, {Fedele},
  {Podio}, {Damiani}, {McGehee}, {Street}, {Gizis}, {Sacco}, {Magrini},
  {Flaccomio}, {Orlando}, {Miceli}, {Stelzer}, {Fuchs}, {Chen}, {Pikuz},
  {Frasca}, {Biazzo}, {Codella}, {Pastorello}, {Alcala'}, {Covino}, {Bianchi},
  \& {Nisini}}]{Bonito:18:YSOLSST}
{Bonito}, R., {Hartigan}, P., {Venuti}, L., {et~al.} 2018, arXiv e-prints,
  arXiv:1812.03135.
\newblock \doarXiv{1812.03135}

\bibitem[{Bonito {et~al.}(2021)Bonito, Venuti, Guarcello, Yoachim, Prisinzano,
  Stassun, Giannini, Street, Clarkson, McGehee, Bellm, Gizis, \&
  Hartigan}]{Bonito2021}
Bonito, R., Venuti, L., Guarcello, M.~G., {et~al.} 2021, Young stellar objects
  and their variability with Rubin Observatory LSST.
\newblock
  \url{https://docushare.lsst.org/docushare/dsweb/Get/Document-37625/rbonito.pdf}

\bibitem[{{Burke} {et~al.}(2021){Burke}, {Shen}, {Blaes}, {Gammie}, {Horne},
  {Jiang}, {Liu}, {McHardy}, {Morgan}, {Scaringi}, \&
  {Yang}}]{2021Sci...373..789B}
{Burke}, C.~J., {Shen}, Y., {Blaes}, O., {et~al.} 2021, Science, 373, 789,
  \dodoi{10.1126/science.abg9933}

\bibitem[{{Connolly} {et~al.}(2014){Connolly}, {Angeli}, {Chandrasekharan},
  {Claver}, {Cook}, {Ivezic}, {Jones}, {Krughoff}, {Peng}, {Peterson}, {Petry},
  {Rasmussen}, {Ridgway}, {Saha}, {Sembroski}, {vanderPlas}, \&
  {Yoachim}}]{2014SPIE.9150E..14C}
{Connolly}, A.~J., {Angeli}, G.~Z., {Chandrasekharan}, S., {et~al.} 2014, in
  Society of Photo-Optical Instrumentation Engineers (SPIE) Conference Series,
  Vol. 9150, Modeling, Systems Engineering, and Project Management for
  Astronomy VI, ed. G.~Z. Angeli \& P.~Dierickx, 14

\bibitem[{{Delgado} \& {Reuter}(2016)}]{2016SPIE.9910E..13D}
{Delgado}, F., \& {Reuter}, M.~A. 2016, in \procspie, Vol. 9910, Observatory
  Operations: Strategies, Processes, and Systems VI, 991013

\bibitem[{{G{\'o}rski} {et~al.}(2005){G{\'o}rski}, {Hivon}, {Banday},
  {Wandelt}, {Hansen}, {Reinecke}, \& {Bartelmann}}]{2005ApJ...622..759G}
{G{\'o}rski}, K.~M., {Hivon}, E., {Banday}, A.~J., {et~al.} 2005, \apj, 622,
  759, \dodoi{10.1086/427976}

\bibitem[{{Graham} {et~al.}(2019){Graham}, {Kulkarni}, {Bellm}, {Adams},
  {Barbarino}, {Blagorodnova}, {Bodewits}, {Bolin}, {Brady}, {Cenko}, {Chang},
  {Coughlin}, {De}, {Eadie}, {Farnham}, {Feindt}, {Franckowiak}, {Fremling},
  {Gezari}, {Ghosh}, {Goldstein}, {Golkhou}, {Goobar}, {Ho}, {Huppenkothen},
  {Ivezi{\'c}}, {Jones}, {Juric}, {Kaplan}, {Kasliwal}, {Kelley}, {Kupfer},
  {Lee}, {Lin}, {Lunnan}, {Mahabal}, {Miller}, {Ngeow}, {Nugent}, {Ofek},
  {Prince}, {Rauch}, {van Roestel}, {Schulze}, {Singer}, {Sollerman}, {Taddia},
  {Yan}, {Ye}, {Yu}, {Barlow}, {Bauer}, {Beck}, {Belicki}, {Biswas}, {Brinnel},
  {Brooke}, {Bue}, {Bulla}, {Burruss}, {Connolly}, {Cromer}, {Cunningham},
  {Dekany}, {Delacroix}, {Desai}, {Duev}, {Feeney}, {Flynn}, {Frederick},
  {Gal-Yam}, {Giomi}, {Groom}, {Hacopians}, {Hale}, {Helou}, {Henning},
  {Hover}, {Hillenbrand}, {Howell}, {Hung}, {Imel}, {Ip}, {Jackson}, {Kaspi},
  {Kaye}, {Kowalski}, {Kramer}, {Kuhn}, {Landry}, {Laher}, {Mao}, {Masci},
  {Monkewitz}, {Murphy}, {Nordin}, {Patterson}, {Penprase}, {Porter},
  {Rebbapragada}, {Reiley}, {Riddle}, {Rigault}, {Rodriguez}, {Rusholme}, {van
  Santen}, {Shupe}, {Smith}, {Soumagnac}, {Stein}, {Surace}, {Szkody}, {Terek},
  {Van Sistine}, {van Velzen}, {Vestrand}, {Walters}, {Ward}, {Zhang}, \&
  {Zolkower}}]{Graham:19:ZTFScience}
{Graham}, M.~J., {Kulkarni}, S.~R., {Bellm}, E.~C., {et~al.} 2019, \pasp, 131,
  078001

\bibitem[{Harris {et~al.}(2020)Harris, Millman, van~der Walt, Gommers,
  Virtanen, Cournapeau, Wieser, Taylor, Berg, Smith, Kern, Picus, Hoyer, van
  Kerkwijk, Brett, Haldane, del R{\'{i}}o, Wiebe, Peterson,
  G{\'{e}}rard-Marchant, Sheppard, Reddy, Weckesser, Abbasi, Gohlke, \&
  Oliphant}]{harris2020array}
Harris, C.~R., Millman, K.~J., van~der Walt, S.~J., {et~al.} 2020, Nature, 585,
  357, \dodoi{10.1038/s41586-020-2649-2}

\bibitem[{Hunter(2007)}]{Hunter:2007}
Hunter, J.~D. 2007, Computing In Science \& Engineering, 9, 90

\bibitem[{{Ivezi{\'c}} {et~al.}(2019){Ivezi{\'c}}, {Kahn}, {Tyson}, {Abel},
  {Acosta}, {Allsman}, {Alonso}, {AlSayyad}, {Anderson}, {Andrew}, {Angel},
  {Angeli}, {Ansari}, {Antilogus}, {Araujo}, {Armstrong}, {Arndt}, {Astier},
  {Aubourg}, {Auza}, {Axelrod}, {Bard}, {Barr}, {Barrau}, {Bartlett}, {Bauer},
  {Bauman}, {Baumont}, {Bechtol}, {Bechtol}, {Becker}, {Becla}, {Beldica},
  {Bellavia}, {Bianco}, {Biswas}, {Blanc}, {Blazek}, {Blandford}, {Bloom},
  {Bogart}, {Bond}, {Booth}, {Borgland}, {Borne}, {Bosch}, {Boutigny},
  {Brackett}, {Bradshaw}, {Brandt}, {Brown}, {Bullock}, {Burchat}, {Burke},
  {Cagnoli}, {Calabrese}, {Callahan}, {Callen}, {Carlin}, {Carlson},
  {Chandrasekharan}, {Charles-Emerson}, {Chesley}, {Cheu}, {Chiang}, {Chiang},
  {Chirino}, {Chow}, {Ciardi}, {Claver}, {Cohen-Tanugi}, {Cockrum}, {Coles},
  {Connolly}, {Cook}, {Cooray}, {Covey}, {Cribbs}, {Cui}, {Cutri}, {Daly},
  {Daniel}, {Daruich}, {Daubard}, {Daues}, {Dawson}, {Delgado}, {Dellapenna},
  {de Peyster}, {de Val-Borro}, {Digel}, {Doherty}, {Dubois},
  {Dubois-Felsmann}, {Durech}, {Economou}, {Eifler}, {Eracleous}, {Emmons},
  {Fausti Neto}, {Ferguson}, {Figueroa}, {Fisher-Levine}, {Focke}, {Foss},
  {Frank}, {Freemon}, {Gangler}, {Gawiser}, {Geary}, {Gee}, {Geha}, {Gessner},
  {Gibson}, {Gilmore}, {Glanzman}, {Glick}, {Goldina}, {Goldstein}, {Goodenow},
  {Graham}, {Gressler}, {Gris}, {Guy}, {Guyonnet}, {Haller}, {Harris},
  {Hascall}, {Haupt}, {Hernandez}, {Herrmann}, {Hileman}, {Hoblitt}, {Hodgson},
  {Hogan}, {Howard}, {Huang}, {Huffer}, {Ingraham}, {Innes}, {Jacoby}, {Jain},
  {Jammes}, {Jee}, {Jenness}, {Jernigan}, {Jevremovi{\'c}}, {Johns}, {Johnson},
  {Johnson}, {Jones}, {Juramy-Gilles}, {Juri{\'c}}, {Kalirai}, {Kallivayalil},
  {Kalmbach}, {Kantor}, {Karst}, {Kasliwal}, {Kelly}, {Kessler}, {Kinnison},
  {Kirkby}, {Knox}, {Kotov}, {Krabbendam}, {Krughoff}, {Kub{\'a}nek},
  {Kuczewski}, {Kulkarni}, {Ku}, {Kurita}, {Lage}, {Lambert}, {Lange},
  {Langton}, {Le Guillou}, {Levine}, {Liang}, {Lim}, {Lintott}, {Long},
  {Lopez}, {Lotz}, {Lupton}, {Lust}, {MacArthur}, {Mahabal}, {Mandelbaum},
  {Markiewicz}, {Marsh}, {Marshall}, {Marshall}, {May}, {McKercher}, {McQueen},
  {Meyers}, {Migliore}, {Miller}, {Mills}, {Miraval}, {Moeyens}, {Moolekamp},
  {Monet}, {Moniez}, {Monkewitz}, {Montgomery}, {Morrison}, {Mueller},
  {Muller}, {Mu{\~n}oz Arancibia}, {Neill}, {Newbry}, {Nief}, {Nomerotski},
  {Nordby}, {O'Connor}, {Oliver}, {Olivier}, {Olsen}, {O'Mullane}, {Ortiz},
  {Osier}, {Owen}, {Pain}, {Palecek}, {Parejko}, {Parsons}, {Pease},
  {Peterson}, {Peterson}, {Petravick}, {Libby Petrick}, {Petry},
  {Pierfederici}, {Pietrowicz}, {Pike}, {Pinto}, {Plante}, {Plate}, {Plutchak},
  {Price}, {Prouza}, {Radeka}, {Rajagopal}, {Rasmussen}, {Regnault}, {Reil},
  {Reiss}, {Reuter}, {Ridgway}, {Riot}, {Ritz}, {Robinson}, {Roby}, {Roodman},
  {Rosing}, {Roucelle}, {Rumore}, {Russo}, {Saha}, {Sassolas}, {Schalk},
  {Schellart}, {Schindler}, {Schmidt}, {Schneider}, {Schneider}, {Schoening},
  {Schumacher}, {Schwamb}, {Sebag}, {Selvy}, {Sembroski}, {Seppala}, {Serio},
  {Serrano}, {Shaw}, {Shipsey}, {Sick}, {Silvestri}, {Slater}, {Smith},
  {Smith}, {Sobhani}, {Soldahl}, {Storrie-Lombardi}, {Stover}, {Strauss},
  {Street}, {Stubbs}, {Sullivan}, {Sweeney}, {Swinbank}, {Szalay}, {Takacs},
  {Tether}, {Thaler}, {Thayer}, {Thomas}, {Thornton}, {Thukral}, {Tice},
  {Trilling}, {Turri}, {Van Berg}, {Vanden Berk}, {Vetter}, {Virieux},
  {Vucina}, {Wahl}, {Walkowicz}, {Walsh}, {Walter}, {Wang}, {Wang}, {Warner},
  {Wiecha}, {Willman}, {Winters}, {Wittman}, {Wolff}, {Wood-Vasey}, {Wu},
  {Xin}, {Yoachim}, \& {Zhan}}]{2019ApJ...873..111I}
{Ivezi{\'c}}, {\v{Z}}., {Kahn}, S.~M., {Tyson}, J.~A., {et~al.} 2019, \apj,
  873, 111, \dodoi{10.3847/1538-4357/ab042c}

\bibitem[{Jones {et~al.}(2020)Jones, Yoachim, Ivezic, Neilsen, \&
  Ribeiro}]{jones_r_lynne_2020_4048838}
Jones, R.~L., Yoachim, P., Ivezic, Z., Neilsen, E.~H., \& Ribeiro, T. 2020,
  {Survey Strategy and Cadence Choices for the Vera C. Rubin Observatory Legacy
  Survey of Space and Time (LSST)}, v1.2,  Zenodo,
  \dodoi{10.5281/zenodo.4048838}.
\newblock \url{https://doi.org/10.5281/zenodo.4048838}

\bibitem[{{Jones} {et~al.}(2014){Jones}, {Yoachim}, {Chandrasekharan},
  {Connolly}, {Cook}, {Ivezic}, {Krughoff}, {Petry}, \&
  {Ridgway}}]{2014SPIE.9149E..0BJ}
{Jones}, R.~L., {Yoachim}, P., {Chandrasekharan}, S., {et~al.} 2014, in Society
  of Photo-Optical Instrumentation Engineers (SPIE) Conference Series, Vol.
  9149, Observatory Operations: Strategies, Processes, and Systems V, ed. A.~B.
  Peck, C.~R. Benn, \& R.~L. Seaman, 0

\bibitem[{{Jones} {et~al.}(2018){Jones}, {Slater}, {Moeyens}, {Allen},
  {Axelrod}, {Cook}, {Ivezi{\'c}}, {Juri{\'c}}, {Myers}, \&
  {Petry}}]{2018Icar..303..181J}
{Jones}, R.~L., {Slater}, C.~T., {Moeyens}, J., {et~al.} 2018, \icarus, 303,
  181, \dodoi{10.1016/j.icarus.2017.11.033}

\bibitem[{{Moeyens} {et~al.}(2021){Moeyens}, {Juric}, {Ford}, {Bektesevic},
  {Connolly}, {Eggl}, {Ivezi{\'c}}, {Jones}, {Bryce Kalmbach}, \&
  {Smotherman}}]{tmp_Moeyens:21:THOR}
{Moeyens}, J., {Juric}, M., {Ford}, J., {et~al.} 2021, arXiv e-prints,
  arXiv:2105.01056.
\newblock \doarXiv{2105.01056}

\bibitem[{{Naghib} {et~al.}(2019){Naghib}, {Yoachim}, {Vanderbei}, {Connolly},
  \& {Jones}}]{2019AJ....157..151N}
{Naghib}, E., {Yoachim}, P., {Vanderbei}, R.~J., {Connolly}, A.~J., \& {Jones},
  R.~L. 2019, \aj, 157, 151, \dodoi{10.3847/1538-3881/aafece}

\bibitem[{{Raiteri} {et~al.}(2021{\natexlab{a}}){Raiteri}, {Villata},
  {Larionov}, {Jorstad}, {Marscher}, {Weaver}, {Acosta-Pulido}, {Agudo},
  {Andreeva}, {Arkharov}, {Bachev}, {Ben{\'\i}tez}, {Berton}, {Bj{\"o}rklund},
  {Borman}, {Bozhilov}, {Carnerero}, {Carosati}, {Casadio}, {Chen},
  {Damljanovic}, {D'Ammando}, {Escudero}, {Fuentes}, {Giroletti}, {Grishina},
  {Gupta}, {Hagen-Thorn}, {Hart}, {Hiriart}, {Hou}, {Ivanov}, {Kim},
  {Kimeridze}, {Konstantopoulou}, {Kopatskaya}, {Kurtanidze}, {Kurtanidze},
  {L{\"a}hteenm{\"a}ki}, {Larionova}, {Larionova}, {Marchili}, {Markovic},
  {Minev}, {Morozova}, {Myserlis}, {Nakamura}, {Nikiforova}, {Nikolashvili},
  {Otero-Santos}, {Ovcharov}, {Pursimo}, {Rahimov}, {Righini}, {Sakamoto},
  {Savchenko}, {Semkov}, {Shakhovskoy}, {Sigua}, {Stojanovic}, {Strigachev},
  {Thum}, {Tornikoski}, {Traianou}, {Troitskaya}, {Troitskiy}, {Tsai},
  {Valcheva}, {Vasilyev}, {Vince}, \& {Zaharieva}}]{Raiteri:21a}
{Raiteri}, C.~M., {Villata}, M., {Larionov}, V.~M., {et~al.}
  2021{\natexlab{a}}, \mnras, 504, 5629, \dodoi{10.1093/mnras/stab1268}

\bibitem[{{Raiteri} {et~al.}(2021{\natexlab{b}}){Raiteri}, {Villata},
  {Carosati}, {Ben{\'\i}tez}, {Kurtanidze}, {Gupta}, {Mirzaqulov}, {D'Ammando},
  {Larionov}, {Pursimo}, {Acosta-Pulido}, {Baida}, {Balmaverde}, {Bonnoli},
  {Borman}, {Carnerero}, {Chen}, {Dhiman}, {Di Maggio}, {Ehgamberdiev},
  {Hiriart}, {Kimeridze}, {Kurtanidze}, {Lin}, {Lopez}, {Marchini},
  {Matsumoto}, {Mujica}, {Nakamura}, {Nikiforova}, {Nikolashvili}, {Okhmat},
  {Otero-Santos}, {Rizzi}, {Sakamoto}, {Semkov}, {Sigua}, {Stiaccini},
  {Troitsky}, {Tsai}, {Vasilyev}, \& {Zhovtan}}]{Raiteri:21b}
{Raiteri}, C.~M., {Villata}, M., {Carosati}, D., {et~al.} 2021{\natexlab{b}},
  \mnras, 501, 1100, \dodoi{10.1093/mnras/staa3561}

\bibitem[{{Reuter} {et~al.}(2016){Reuter}, {Cook}, {Delgado}, {Petry}, \&
  {Ridgway}}]{2016SPIE.9911E..25R}
{Reuter}, M.~A., {Cook}, K.~H., {Delgado}, F., {Petry}, C.~E., \& {Ridgway},
  S.~T. 2016, in \procspie, Vol. 9911, Modeling, Systems Engineering, and
  Project Management for Astronomy VI, 991125

\bibitem[{Richards {et~al.}(2018)Richards, Yu, Brandt, Ni, Peters, Yang, \&
  Bauer}]{Richards2018}
Richards, G., Yu, W., Brandt, W., {et~al.} 2018, Testing of LSST AGN Selection
  Using Rolling Cadences.
\newblock
  \url{https://docushare.lsstcorp.org/docushare/dsweb/Get/Document-30572/richards_agn_rolling_wfd.pdf}

\bibitem[{{Ricker} {et~al.}(2014){Ricker}, {Winn}, {Vanderspek}, {Latham},
  {Bakos}, {Bean}, {Berta-Thompson}, {Brown}, {Buchhave}, {Butler}, {Butler},
  {Chaplin}, {Charbonneau}, {Christensen-Dalsgaard}, {Clampin}, {Deming},
  {Doty}, {De Lee}, {Dressing}, {Dunham}, {Endl}, {Fressin}, {Ge}, {Henning},
  {Holman}, {Howard}, {Ida}, {Jenkins}, {Jernigan}, {Johnson}, {Kaltenegger},
  {Kawai}, {Kjeldsen}, {Laughlin}, {Levine}, {Lin}, {Lissauer}, {MacQueen},
  {Marcy}, {McCullough}, {Morton}, {Narita}, {Paegert}, {Palle}, {Pepe},
  {Pepper}, {Quirrenbach}, {Rinehart}, {Sasselov}, {Sato}, {Seager},
  {Sozzetti}, {Stassun}, {Sullivan}, {Szentgyorgyi}, {Torres}, {Udry}, \&
  {Villasenor}}]{Ricker:14:TESS}
{Ricker}, G.~R., {Winn}, J.~N., {Vanderspek}, R., {et~al.} 2014, in Society of
  Photo-Optical Instrumentation Engineers (SPIE) Conference Series, Vol. 9143,
  Space Telescopes and Instrumentation 2014: Optical, Infrared, and Millimeter
  Wave, ed. J.~{Oschmann}, Jacobus~M., M.~{Clampin}, G.~G. {Fazio}, \& H.~A.
  {MacEwen}, 914320

\bibitem[{{Ridgway} {et~al.}(2014){Ridgway}, {Matheson}, {Mighell}, {Olsen}, \&
  {Howell}}]{Ridgway:14:TransientDiscoveryRate}
{Ridgway}, S.~T., {Matheson}, T., {Mighell}, K.~J., {Olsen}, K.~A., \&
  {Howell}, S.~B. 2014, \apj, 796, 53, \dodoi{10.1088/0004-637X/796/1/53}

\bibitem[{{Van Der Walt} {et~al.}(2011){Van Der Walt}, {Colbert}, \&
  {Varoquaux}}]{numpy:2011}
{Van Der Walt}, S., {Colbert}, S.~C., \& {Varoquaux}, G. 2011, {Computing in
  Science \& Engineering}, 13, 22, \dodoi{10.1109/MCSE.2011.37}

\bibitem[{Zonca {et~al.}(2019)Zonca, Singer, Lenz, Reinecke, Rosset, Hivon, \&
  Gorski}]{Zonca2019}
Zonca, A., Singer, L., Lenz, D., {et~al.} 2019, Journal of Open Source
  Software, 4, 1298, \dodoi{10.21105/joss.01298}

\end{thebibliography}

\end{document}